\documentclass[twocolumn]{aastex62}

\shorttitle{Energy Deposition at the Upper Atmospheres of the Trappist-1 Planets}
\shortauthors{Cohen et al.}

\begin{document}

\title{Energy Dissipation in the Upper Atmospheres of Trappist-1 Planets}

\email{ofer\_cohen@uml.edu}

\author{Ofer Cohen}
\affiliation{Lowell Center for Space Science and Technology, University of Massachusetts Lowell \\
600 Suffolk St., Lowell, MA 01854, USA}
\affiliation{Harvard-Smithsonian Center for Astrophysics,60 Garden St., Cambridge, MA 02138, USA}

\author{Alex Glocer}
\affiliation{NASA/Goddard Space Flight Center, Greenbelt, Maryland, USA}

\author{Cecilia Garraffo}
\affiliation{Harvard-Smithsonian Center for Astrophysics, \\
60 Garden St., Cambridge, Massachusetts, USA}

\author{Jeremy J. Drake}
\affiliation{Harvard-Smithsonian Center for Astrophysics, \\
60 Garden St., Cambridge, Massachusetts, USA}

\author{Jared M. Bell}
\affiliation{National Institute of Aerospace, 100 Exploration Way, Hampton, VA 23666, USA}

\begin{abstract}

We present a method to quantify the upper-limit of the energy transmitted from the intense stellar wind to the upper atmospheres of three of the Trappist-1 planets (e, f, and g). We use a formalism that treats the system as two electromagnetic regions, where the efficiency of the energy transmission between one region (the stellar wind at the planetary orbits) to the other (the planetary ionospheres) depends on the relation between the conductances and impedances of the two regions. Since the energy flux of the stellar wind is very high at these planetary orbits, we find that for the case of high transmission efficiency (when the conductances and impedances are close in magnitude), the energy dissipation in the upper planetary atmospheres is also very large. On average, the Ohmic energy can reach $0.5-1~W/m^2$, about 1\% of the stellar irradiance and 5-15 times the EUV irradiance. Here, using constant values for the ionospheric conductance, we demonstrate that the stellar wind energy could potentially drive large atmospheric heating in terrestrial planets, as well as in hot jupiters. More detailed calculations are needed to assess the ionospheric conductance and to determine more accurately the amount of heating the stellar wind can drive in close-orbit planets.       

\end{abstract}


\section{Introduction} 

The recent discovery of seven Earth-size terrestrial planets in the Trappist-1 system \citep{Gillon.etal:17} has stimulated the possibility of detecting habitable planets in nearby systems. Indeed, three of the seven Trappist-1 planets - Trappist-1e, Trappist-1f, and Trappist-1g - are in the Habitable Zone (HZ) defined as a bounded distance from the host star at which the planetary equilibrium temperature allows water to exist in liquid form on the planetary surface. A growing number of studies have been published on the Trappist-1 system in the short time since its discovery. These include studies of the formation and evolution of the planetary system and its planets \citep[e.g.,][]{Barr:17,Burgasser:17,Luger:17,Ormel:17,Quarles:17,Tamayo:17}, the atmospheres of the Trappist-1 planets \cite[e.g.,][]{Alberti:17,Bourrier:17,Tilley:17,Wolf:17}, and the chance for life to exist on the Trappist-1 planets \citep{Lingam:17}. 

While detections of terrestrial planets in the HZ of Trappist-1 are exciting, a major potential problem for their habitability is the fact that the HZ around faint M-dwarf stars is extremely close to the host star. It may be located at a distance of less than 0.1~AU and essentially inside the stellar corona. Indeed, the orbital distances of Trappist-1 e, f, and g are 0.028~AU, 0.037~AU, and 0.045~AU, respectively. In such close orbits, the conditions of the stellar environment are much more extreme than those experienced by a planet located like the Earth, much further from the host star. These include increased stellar energetic radiation, enhanced density of the stellar wind (resulting in enhanced dynamic pressure), and enhanced magnitude of the stellar wind magnetic field (resulting in enhanced magnetic pressure). These extreme conditions may lead to evaporation and stripping of the planetary atmosphere until they completely lost. Thus, the chance of habitability could be greatly reduced. 

Few generic studies have been performed to estimate the atmospheric loss from close-orbit planets \citep[e.g.,][]{Cohen.etal:15,Airapetian:17,Dong:17b}. Recent studies have estimated the space environment conditions and the atmospheric loss on Trappist-1 \citep{Roettenbacher:17,Garraffo:17} and the recently detected Proxima Centauri b \citep{Garraffo.etal:16,Dong:17a,Garcia-Sage:17}. All these studies have pointed to a very high mass loss rates from these close-in planets, suggesting their atmospheres may be completely eliminated over their lifetimes. We stress that these estimates did not attempt to demonstrate how the atmospheres can be formed, which is itself another theoretical challenge.

One key aspect in estimating the ability of a close-in planet to sustain its atmosphere is to quantify the energy input from the stellar radiation and the stellar wind at the location of the planet. Detailed observations and estimation of the stellar EUV and bolometric luminosity at the orbits of the Trappist-1 planets have recently been obtained by \cite{Wheatley.etal:17}. In the study presented here, we quantify the total energy input carried by the stellar wind in the vicinity of the three potentially habitable Trappist-1 planets, and estimate the amount of energy that is delivered to their upper atmospheres, assuming atmospheres do exist. We use the radiation energies obtained by  \cite{Wheatley.etal:17} as reference for the stellar wind energy. 

It is known from our own solar system, that the solar wind energy is dissipated in upper planetary atmospheres in the form of Joule Heating or Ohmic dissipation in the ionosphere. The ionosphere is the layer of the upper atmosphere at which photoionization creates a peak in the electron density so that conductivity becomes finite. The ionosphere allows field-aligned currents, that flow from the magnetosphere (in the case of magnetized planets) or the induced magnetosphere (in the case of non-magnetized planets), to close through it, while dissipating the energy due to its resistive nature \citep[see e.g.,][for a complete description of the process]{Kivelson.Russell:95,Gombosi:04}. The dissipating energy depends on the solar wind driver, which drives the field-aligned currents, and the conductivity in the ionosphere, which depends on the atmospheric conditions, composition, and ionization. In general, the large-scale, ambient ionospheric conductivity is dominated by the so-called Pedersen conductivity, which is the component of the conductivity tensor responsible for the electric field that is perpendicular to the ambient magnetic field. In the case of the Earth's ionosphere, this is the electric field that is perpendicular to both the solar wind velocity and the solar wind magnetic field, driving a current that flows across the region where the Earth's magnetic field is open to the solar wind. This region of open field lines is called the polar cap and the electric field across it is associated with a Cross Polar Cap Potential (CPCP). 

There is some evidence that during a strong solar wind driver, the CPCP is saturated \citep[see][for longer description with reference therein]{KivelsonRidley:08}. In particular, the saturation might occur when the solar wind Alfv\'enic Mach number is very low, even below one. In that case, the interaction of the moving body (magnetized or non-magnetized) with the sub-Alfv\'enic flow generates the topology of {\it Alfv\'en Wings} - two standing lobes expanding an angle that depends on the velocity of the body and the Alfv\'en speed \citep{Drell65,Neubauer80,Neubauer98}. The energy transfer from the stellar wind to the ionosphere during such conditions can be treated in an idealized wave transmission manner. \cite{KivelsonRidley:08} (KR08 hereafter) have treated the ionosphere as a spherical conductor with finite conductivity, and the incoming solar wind as an electromagnetic wave. They showed that the energy transmitted from the solar wind to the ionosphere can be estimated as the transmitted energy of the incoming electromagnetic wave. 

Here, we adopt the formalism of KR08 to estimate the energy input from the extreme stellar wind of the Trappist-1 planets onto the atmospheres of the e, f, and g planets. We describe the formalism in Section~\ref{Fromalism} and present the results in Section~\ref{Results}. We discuss the consequences of the results for the atmospheres of Trappist-1 in Section~\ref{Discussion} and conclude our findings in Section~\ref{Conclusions}.


\section{Wave Transmission Formalism of the Stellar Wind Energy Input}
\label{Fromalism}

We now review the method which is described in KR08 in the context of exoplanets. The stellar wind at the vicinity of a planet has a velocity $\mathbf{v}_{sw}$ and a magnetic field $\mathbf{B}_{sw}$. Therefore, the motional electric field, $\mathbf{E}_{sw}$ can be obtain as:

\begin{equation}
\mathbf{E}_{sw}=-\mathbf{v}_{sw}\times \mathbf{B}_{sw}, 
\end{equation} 
and the magnitude of the electric field is given by:

\begin{equation}
|\mathbf{E}_{sw}|=|\mathbf{v}_{sw}|\cdot |\mathbf{B}_{sw}|. 
\end{equation} 

The local Alfv\'en speed of the stellar wind is given by:

\begin{equation}
v_A=\frac{B_{sw}}{\sqrt{\mu_0 \rho_{sw}}},
\end{equation} 

Thus, we can define a conductivity associated with the incoming stellar wind and its Alfv\'en speed that is given by:

\begin{equation}
\Sigma_A=(\mu_0v_A)^{-1}\;[Siemens],
\end{equation} 
where we can also define an associated Alfv\'enic impedance which is the inverse of the Alfv\'enic conductivity:

\begin{equation}
\Sigma^{-1}_A=\mu_0v_A\;[ohm].
\end{equation} 

The local Pedersen conductivity at certain altitude, $\sigma_P$, is a function of the local electron density, $N_e$, the charge, $e$, the ion and electron masses, $m_i$ and $m_e$, respectively, the ion and electron stress collision frequencies, $\nu_i$ and $\nu_e$, respectively, and the ion and electron plasma frequencies, $\Omega_i$ and $\Omega_e$, respectively \citep{Kivelson.Russell:95}:

\begin{equation}
\sigma_p=e^2N_e \left[ \frac{\nu_i}{m_i \left( \nu^2_i+\Omega^2_i \right)} + \frac{\nu_e}{m_e \left( \nu^2_e+\Omega^2_e \right)}\right]~[S/m], 
\end{equation}

The height-integrated Pedersen conductance, $\Sigma_P$ is the column height integral of $\sigma_P$, where we can introduce an associated Pedersen impedance simply defined as $\Sigma^{-1}_P$. Typical Earth values for the height integrated conductance are 1-10 \citep[e.g.,][]{KivelsonRidley:08}. Here we test our calculations against assumed values of $\Sigma_p=0.1,1,5,10,50,~and~100~[S]$.

Assuming the stellar wind electric field can be considered as an electromagnetic wave, we can use the impedances defined above to calculate the reflection and transmission of the incoming electric field wave, $|E_i|$. The reflected electric field is given by:

\begin{equation}
|E_r|=|E_i| \frac{\left( \Sigma^{-1}_P-\Sigma^{-1}_A \right)}{\left( \Sigma^{-1}_P+\Sigma^{-1}_A \right)},
\end{equation} 
Note that when $\Sigma^{-1}_P$ is smaller than $\Sigma^{-1}_A$, the reflected wave has an opposite sign to that of the incoming wave. Thus, the transmitted electric field is given by:

\begin{equation}
|E_t|=|E_r|+|E_i|=2|\mathbf{E}_{sw}| \frac{\Sigma^{-1}_P}{\left( \Sigma^{-1}_P+\Sigma^{-1}_A \right)}
\end{equation} 
where the transmission of the electric field is expected to be most significant where the Alfv\'enic and Pedersen conductances are close in magnitude. 

Once the transmitted electric field is obtained, we can estimate the energy flux that is dissipated in the planetary ionosphere via Ohmic dissipation, $Q_t$, as:

\begin{equation}
\label{Qt}
Q_t=\frac{j^2}{\Sigma_P}=\frac{\Sigma_P^2|E_t|^2}{\Sigma_P}=\Sigma_P|E_t|^2~[W/m^2].
\end{equation} 
Note that since we use the height integrated conductance, the units of $Q_t$ are not of $\mathbf{J}\cdot\mathbf{E}$ but of $\mathbf{J}\cdot\mathbf{E}$ multiplied by a length scale, which gives energy flux.


\section{Results}
\label{Results}

\begin{figure*}[h!]
\centering
\includegraphics[width=6.in]{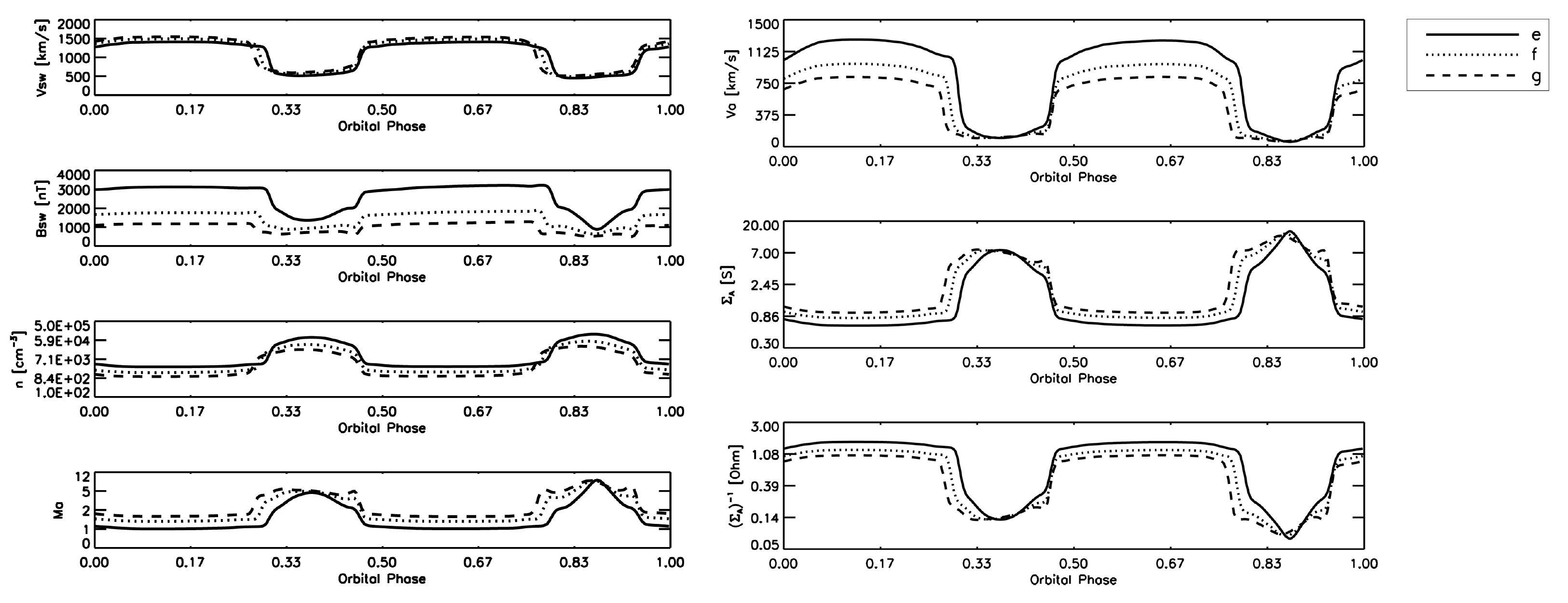}
\caption{Left: stellar wind speed (top), magnetic field (second), number density (third), and Alfv\'enic Mach number (bottom) as a function of orbital phase of Trappist-1 e, f, and g \citep[taken from][]{Garraffo:17}. Right: Alfv\'enic velocity (top), conductance (middle), and impedance (bottom) along the orbits of the three planets.}
\label{fig:f1}
\end{figure*}

Figure~\ref{fig:f1} shows the stellar wind parameters along the orbits of Trappist-1 e, f, and g. The parameters were obtained from the MHD wind simulation presented in \cite{Garraffo:17} (case with an average field of 600G). The stellar wind parameters are more extreme than typical solar wind conditions near the Earth, with wind speeds of 1.5-2 times that of the solar wind, magnetic field 100-1000 times larger than the solar wind magnetic field, and 100-1000 times more dense wind compared to the solar wind near the Earth. It can be seen that during most of the orbit, the three planets reside in a low-Alfv\'enic Mach number (less than 2), where Trappist-1e experiences a plasma environment with $M_A\approx 1$ for most of its orbit. The Alfv\'en Mach number increases when the planets cross the more dense streamer regions, near the orbital phases of 0.35 and 0.9. 

Figure~\ref{fig:f1} also shows the value of the Alfv\'en speed, $V_A$, the Alfv\'enic conductivity, $\Sigma_A$, and the Alfv\'enic impedance, $\Sigma^{-1}_A$, as a function of the orbital phase of Trappist-1 e, f, and g. The conductance is below 1 for most of the orbit but reaches values of 7-10 during the streamer crossings. The impedance values are around 1 most of the orbit but become about 10 times smaller during the streamer crossings.

\begin{figure*}[h!]
\centering
\includegraphics[width=6.in]{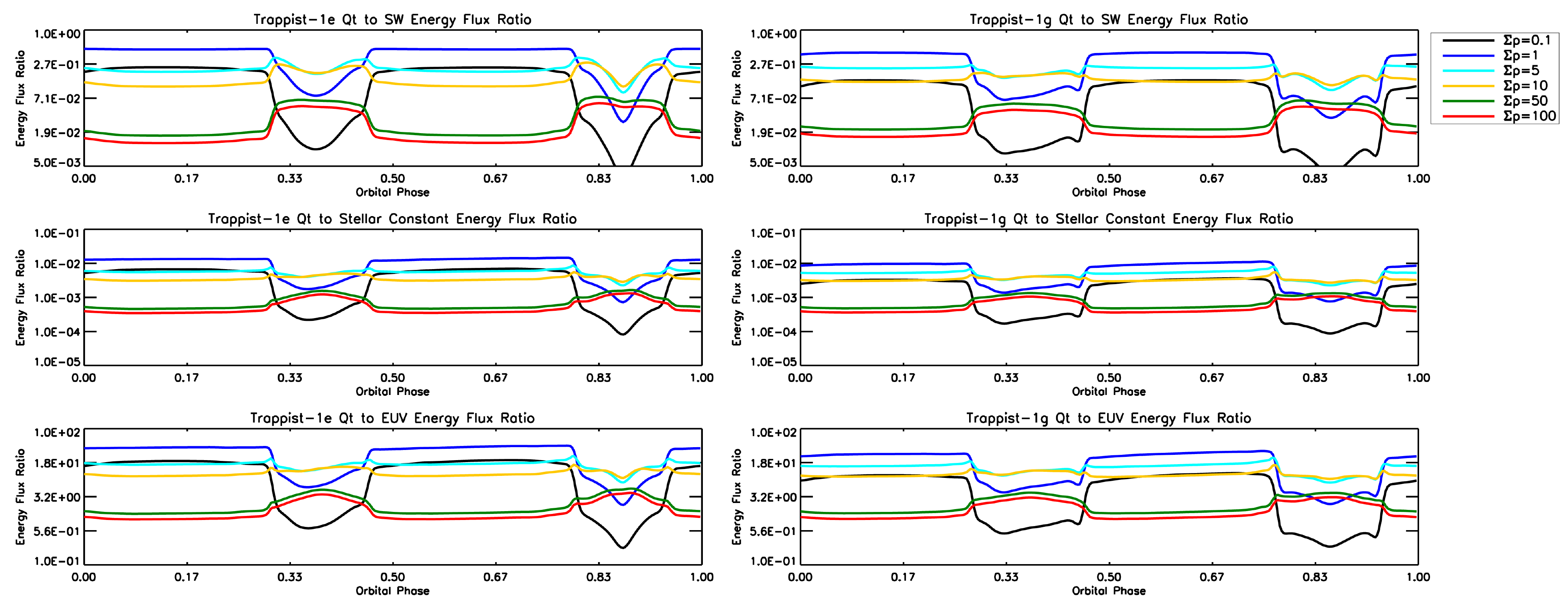}
\caption{Ratio of the transmitted energy flux, $Q_t$ and the total stellar wind energy input (top), the total stellar radiation flux (middle), and the EUV radiation flux (bottom) as a function of the orbital phase for different values of $\Sigma_P$. Results are shown for Trappist-1e (left) and Trappist-1g (right). The results for Trappist-1f show an intermittent behavior and thus are not shown here.}
\label{fig:f2}
\end{figure*}

Figure~\ref{fig:f2} shows the transmitted energy deposited into the planetary atmospheres of Trappist-1 e and g as a function of orbital phase. The results for Trappist-1f lie in between these two cases and therefore are not shown here. We normalize the energy flux to three values: i) the energy flux of the stellar wind; ii) the stellar irradiance; and iii) the stellar EUV irradiance. We estimate the stellar wind energy flux, in $W/m^2$, as the sum of the dynamic ($\rho_{sw} v^2_{sw}$) and magnetic ($B^2_{sw}/8\pi$) pressures multiplied by the wind velocity:

\begin{equation}
F_{sw}=\left( \rho_{sw} v^2_{sw} + B^2_{sw}/8\pi\right)\cdot v_{sw}
\end{equation} 

The results show that the transmitted energy is significant for values of $\Sigma_P<10$ during the orbital phases where the planet resides in a low Alfv\'enic Mach number. Higher values of the Pedersen conductance, or orbital phases at which the Alfv\'enic Mach number is higher than 2, seem insufficient to enable deposition of a significant amount of heating in the upper atmosphere from the intense stellar wind due to the reflection of most of the Alfv\'enic energy input. As expected, the energy transmission is most efficient when the values of the stellar wind and ionospheric conductances are close to each other. In these cases, about 10-50\% of the stellar wind input energy is transmitted with $\Sigma_P=0.1,1,~and~5$ for Trappist-1e, with similar or slightly lower values for Trappist-1 f, and g. This transmitted energy translates to about 0.5-1\% of the stellar irradiance, and 5-15 times the EUV energy flux. 

Table~\ref{table:t1} shows the stellar radiation fluxes \cite[taken and extrapolated from][]{Wheatley.etal:17}. It also shows the orbital-averaged stellar wind input parameters and the heat fluxes for $\Sigma_P=1~and~10$ for the three planets. Note that these avreraged values do not exactly follow Eq.~\ref{Qt}


\section{Discussion}
\label{Discussion}

We quantify the energy deposition from the extreme stellar wind of Trappist-1 into its planets' atmospheres. We use the formalism from KR08 to relate the impedances associated with the stellar wind and the planetary ionosphere to the energy deposition. In a way, this formalism provides an efficiency of the energy transfer from the stellar wind driver to the conducting layer (the ionosphere) in a generalized electromagnetic energy transmission manner. The efficiency depends on the relationship between the ability of the two mediums to allow or suppress electric currents from flowing in them (i.e., the conductances and impedances). 

It is important to note that here we compare the energy inputs to the planet in terms of energy flux, and that in the KV08 formalism, the stellar wind energy flux is assumed to be transmitted in the area covered by the region where planetary field lines are open to the stellar wind (the polar caps). This area depends on the planetary field strength which is unknown. \cite{Garraffo:17} suggested that in the Trappist-1 planets, all planetary field lines are open to the stellar wind so that the polar cap covers the whole planet. Therefore, it is possible that the energy transmission covers a significant area of the planet. Additionally, we assume here that the angle between the stellar wind magnetic field and the stellar wind velocity is 90$^\circ$. At the Earth, the angle between the two is determined by the Parker Spiral and it is about 45$^\circ$. In the case of the Trappist-1 planets, the two are more or less radial, but an angle of 5-30$^\circ$ between the two vectors could appear due to the fast planetary orbital motion (about $100~km~s^{-1}$). A larger angle is also possible due to the fact that the planets may reside in the sub-Alfv\'enic regime where the wind and magnetic field might not be fully coupled. Taking these two points into account, we offer here an upper-limit for the stellar wind available energy and its transmission to the upper atmosphere, where even 10\% of this energy is still very high.

We find that for the cases where the two impedances are close in magnitude, the efficiency of the energy transmission from the wind to the ionosphere is high. Since the stellar wind energy flux is very large, the dissipated energy flux is also very large - 0.5-1\% of the total stellar irradiance and 5-15 times higher than the EUV irradiance. We conclude that the upper atmospheres of close-orbit planets, such as the Trappist-1 planets, could suffer from an intense Ohmic heating sourced in the intense stellar wind input using constant values for the ionospheric conductance. However, it is not trivial to estimate how $\Sigma_P$ changes with the intense EUV radiation in close-in planets. On one hand, the increased ionization should push the ionosphere down to regions where the electron density is higher. On the other hand, this will increase the collision frequencies. Therefore, a more detailed calculation of the integrated ionospheric conductance is needed using Ionosphere-Thermosphere models \citep[e.g.,][]{Ridley:07,Deng:11,Bell14}.

Our estimates are also relevant to the problem of hot-jupiter inflation, which requires additional heating to explain the observed inflation in this planet population. It has been suggested that Ohmic dissipation can be driven by the strong zonal winds in tidally-locked planets and the planetary magnetic field \citep[e.g.,][]{Batygin:10,Rauscher:13}. \citet{Rogers:14a,Rogers:14b} have shown using a full MHD model that such Ohmic dissipation is possible, but it fragments and cannot support the necessary heating. \cite{Koskinen:14} have shown that sufficient dissipation can only occur in the upper parts of the atmosphere, above the 10 mbar level. Our work here demonstrates the potential of Ohmic dissipation in the ionosphere, driven by the intense stellar wind, to provide additional heating. Such a mechanism requires investigation beyond the scope of this paper. 


\section{Conclusions}
\label{Conclusions}

We adopt a method to quantify the energy transfer and efficiency from the solar wind to the Earth's ionosphere to three of the close-orbit planets orbiting Trappist-1 in order to estimate the order of magnitude of the Ohmic heating in the planets' ionospheres. The method relates the conductances and impedances of the stellar wind and the ionosphere to calculate the amount of energy transmitted into the ionosphere in the form of Ohmic dissipation. We use an assumed set of ionospheric conductances and find that for values that are less than $10~S$, the dissipated energy can reach $0.5-1~W/m^2$, which can drive large continuous heating in the upper atmospheres of exoplanets, and potentially take part in hot jupiter inflation. We conclude that further modeling of the ionospheric conductance under the extreme conditions of close-orbits planets is needed to better estimate the dissipated energy.


\acknowledgments

We thank the unknown referee for her/his useful comments and suggestions. We also thank Margaret Kivelson and Aaron Ridley for their useful input. The work presented here is part of NASA NExSS The Living Breathing Planet project supported by NASA Astrobiology Institute grant NNX15AE05G. 




\begin{table*}[h!]
\caption {Stellar Fluxes and Average Orbital Values for the Trappist-1 Planets} 
\begin{center}
\begin{tabular}{  p{0.5in}  p{0.5in} p{0.5in}  p{0.5in} p{0.5in}  p{0.5in}  p{0.5in}  p{0.5in}  p{0.5in}  p{0.5in} }
\hline
{\bf Planet Name}&{\bf Semi-major Axis $[AU]$} &{\bf Stellar Constant $[W~m^{-2}]$} & {\bf EUV Flux  $[W~m^{-2}]$}&$B_{sw}~[nT]$&$v_A~[km/s]$&$E_{sw}~[V/m]$ &Wind Energy Flux $[W/m^{2}]$&$Q_t(\Sigma_P=1)~[W/m^2]$&$Q_t(\Sigma_P=10)~[W/m^2]$\\
\hline
Trappist-1e&0.028&867&0.30&2641&842&3.15&50&8.71&3.16\\
Trappist-1f&0.037&496&0.17&1480&636&1.85&30&4.00&1.75\\
Trappist-1g&0.045&335&0.12&989&527&1.26&20&2.24&1.13\\
\hline
\end{tabular}
\end{center}
\label{table:t1}
\end{table*}


\end{document}